\documentclass[aps,prl,twocolumn,groupedaddress]{revtex4}
\usepackage{graphicx}
\usepackage{amsmath}
\usepackage{amsbsy}

\begin{document}

\title{Can an evolving Universe host a static event horizon?}
\author{Aharon Davidson}
\email[Email: ]{davidson@bgu.ac.il}
\affiliation{Physics Department, Ben-Gurion University of the Negev,
Beer-Sheva 84105, Israel}
\author{Shimon Rubin}
\email[Email: ]{rubin.shim@gmail.com}
\affiliation{Physics Department, Ben-Gurion University of the Negev,
Beer-Sheva 84105, Israel}
\author{Yosef Verbin}
\email[Email: ]{verbin@openu.ac.il}
\affiliation{Department of Natural Sciences, The Open University of Israel,
Raanana 43107, Israel}

\date{November 1, 2012}

\begin{abstract}
	We prove the existence of general relativistic perfect fluid black
	hole solutions, and demonstrate the phenomenon for the
	$P=w\rho$ class of equations of state.
	While admitting a local time-like Killing vector on the event horizon
	itself, the various black hole configurations are necessarily time
	dependent (thereby avoiding a well known no-go theorem) away
	from the horizon.
	Consistently, Hawking's imaginary time periodicity is globally
	manifest on the entire spacetime manifold.
\end{abstract}

\pacs{}
\maketitle

\section{Introduction}

There is by now abundant evidence that our Universe is expanding on average.
Nevertheless, much of the works on black holes, which play a central role in modern astrophysics,
focus on stationary and asymptotically flat situation. It is therefore desirable to have black hole
models embedded in cosmological environment to see if those objects can reveal some unexpected
features of the underlying theory of gravity.
Historically, an investigation of the effects of the cosmological expansion on local systems, was motivated
by the question of whether an atom, a star, Solar System, galaxy or any other bounded system expand
following the rest of the universe. Although this question has a long history dating back to the 1933 paper
by McVittie \cite{McVittie1} introducing a spacetime metric that represents a point mass embedded in a
Friedmann-Lema\^itre-Robertson-Walker (FLRW) universe, one still lacks an affirmative answer to this
open problem in general relativity.
In fact, the physical properties of the McVittie solution are an active field
of research and discussed by several authors, e.g., \cite{Nolan}, \cite{Kaloper}, \cite{Carrera}, \cite{Faraoni}, \cite{LakeAbdelqader}.
Later work by Einstein and Straus \cite{Einstein} describes a patchwork
of Schwarzschild black holes with an FLRW universe, while subsequent generalizations of their ideas \cite{BonnorPetrosian}
replace the external FLRW metric by the inhomogeneous Lema\^itre-Tolman-Bondi (LTB) dust time-dependent
solution \cite{LTB} and by its non spherically-symmetric generalization \cite{Szekeres}.
The former, however, are time symmetric black hole solutions and so they do not describe at a satisfactory
level any phenomenon associated with a dynamical black hole in an evolving Universe. Such processes are
expected to play a significant role in black hole formation, gravitational collapse, evolution of primordial black holes, etc.
This has invoked numerous studies, e.g.,  \cite{Vaidiya}, \cite{SultanaDyer}, \cite{Gao}, \cite{Maeda}, \cite{Nandra}
where new non-static solutions have been considered.
Overall the references to the inhomogeneous solutions in general relativity are quite extensive, e.g.,
see \cite{Krasinski1}, \cite{Krasinski2} and references within.
In this work we are mainly motivated by the question if an evolving and non-stationary universe can host
a static event horizon.

The event horizon is a central object in the description of a black hole, with the Schwarzschild solution being
its most prominent representative. The near horizon structure of the latter is a product of a Rindler space and a
two sphere, which we write as $\textit{Rindler}\times\text{S}^{2}$, and defines the corresponding causal
structure. Specifically, the Rindler piece in the metric typically characterizes the event horizon as a boundary
of region in spacetime from behind which no causal signals can reach the observers sitting far away at infinity.
The Euclidean Rindler metric that is obtained by substituting $t=-it_{E}$ yields flat two-dimensional Euclidean
metric written in polar coordinates, provided the angular variable has the correct periodicity. If the periodicity
is different, then the geometry would have a conical singularity at the would have been horizon $x=0$. Finally, the relation
$\beta=\hbar\Delta\tau$ between the periodicity $\Delta\tau$ in Euclidean time $t_{E}$, and the inverse temperature
$\beta$, leads towards the celebrated Hawking-Bekenstein black hole temperature. Even though we discussed
the temperature in the context of vacuum
Schwarzschild black hole, the concept of black hole temperature that owes its life to the "no conical  singularity" demand, is true even in a presence of matter because the near horizon geometry is always Rindler-like. In this
work we take one step further and assume that a horizon and its underlying Rindler-like structure in the
near-horizon limit, also exists in the time dependent and spherically symmetric case. In particular we will assume
that on the horizon itself the metric is static, i.e. we postulate the existence of a local time-like Killing field.
This in contrast to the Schwarzschild, or Schwarzschild-deSitter black holes where time-like Killing fields are
manifest in any point.

The rest of this work is organized as follows. First, we shortly discuss the McVittie solution, analyze the general structure
of the perfect fluid energy-momentum tensor, the emerging baryon number conservation
and the Euler relativistic equations. Then we introduce the perfect fluid isotropic condition and
the perfect fluid equation of state which constitute our basic equations for the corresponding spherically
symmetric metric components. Their solution yields the corresponding spherically symmetric configuration in
the presence of a perfect fluid. We solve those equations perturbatively, order by order, with dimensionless
radial coordinate serving as an expansion parameter, around the local Killing horizon.
The Rindler metric on this static surface serves as boundary condition in our iterative series solution,
and defines the corresponding causal structure.

\subsection{The McVittie solution}

One of the most prominent representatives of non-homogeneous and non-stationary exact solutions of Einstein field
equations (adopting from this point onward the natural units  $16\pi G_{N}=c=1$)
\begin{equation}
          \mathcal{R}_{\mu \nu}-\frac{1}{2}g_{\mu \nu}\mathcal{R}=-\frac{1}{2}\mathcal{T}_{\mu \nu}
          \label{Einstein}
\end{equation}
is the McVittie line element, in its special $k=0$ case, explicitly given by
\begin{equation}
       ds^{2}=-\left( \frac{1-\mu }{1+\mu }\right) ^{2}dt^{2}+\left( 1+\mu \right)
      ^{4}a\left(t\right)^{2}\left( dr^{2}+r^{2}d\Omega^{2}\right).
      \label{McVittie}
\end{equation}
Here, $\mu$ is a dimensionless quantity defined by
\begin{equation}
       \mu =\frac{m}{2a\left( t\right)r},
\end{equation}
$r$ is the radial coordinate and $d\Omega^{2}=d\theta^{2}+\sin(\theta)^{2}d \phi^{2}$ is the solid angle on a 2-sphere.
The corresponding metric is an exact solution of Einstein field equations
for any constant $m$, provided the corresponding energy density is a function
of the time coordinate alone. It approaches the FLRW universe in the $\mu \ll 1$ limit,
and tends to the Schwarzschild solution in the static case. Although admitting those
two tenable limit cases, it does not cover other cases of interest in the interface between
cosmology and local black hole physics, and cannot be regarded as general problem solution
for the following reasons:

(i) Although McVittie's solution tends to the pure FLRW solution in the far region, it falls
short to match the standard description in terms of a perfect fluid in the near region.
Specifically, any barotropic equation of state $P=P(\rho)$ where $\rho $ and $P$ are the energy density
and isotropic pressure as measured in the instantaneous rest frame of the fluid at the corresponding point,
is incompatible with this solution.
In particular, assuming that the underlying matter is described by the perfect fluid energy-momentum
tensor given by
\begin{equation}
       \mathcal{T}^{\mu \nu }=-\left( \rho +P\right) U^{\mu }U^{\nu }-Pg^{\mu \nu }
       \label{PFEMT}
\end{equation}
 where $U^{\mu}$ is the four
velocity vector that satisfies the normalization condition
\begin{equation}
       U^{\mu }U_{\mu }=-1,
       \label{4vel}
\end{equation}
one is led towards $\rho$ which is a function of time $t$, implying that there is no $P=P(\rho)$ equation of state.

(ii) McVittie's line element is over restrictive since it prevents the hole from accreting, and enforces
its energy density to be exactly homogeneous on some set of spatial slices.

Additional analysis of McVittie solution may be performed by using the concept of the apparent horizon \cite{Hayward}.
The latter defined as a locus of vanishing geodesic expansion, which in the context of the McVittie's metric \cite{Kaloper} yields

\begin{equation}
          1-\frac{2m}{r}-H^{2}r^{2}= 0,
\label{roots}
\end{equation}
which expresses $r$ as a function of $t$ through the time-dependence of the Hubble parameter $H$.
As time goes on the area of the apparent horizon changes, in such a way so that $r_{-}$-the smaller root of Eq.(\ref{roots})
moves inward as $H$ decreases with time towards its asymptotic de Sitter value, while the larger root moves outward.
In the limit $t \rightarrow \infty$ as the Hubble parameter tends to some asymptotic positive value $H \rightarrow H_{0}>0$,
one may show that $r=r_{-}$, $t= \infty$ is a regular event horizon. In the particular $H_{0}=0$ case, on the other hand,
the $r=r_{-}$ surface does not acquire a straightforward meaning as an event horizon, due to
soft singularities of scalars constructed from quantities that involve two derivatives of the Riemann tensor.
In this work we deal with a static event horizon.

To summarize, although the McVittie solution describes a black hole, at least for positive $H_{0}$, it still
does not encompass some effects and properties one would expect for a physical
black hole in a universe full of matter or radiation.

%
%
%
%
%

\subsection{Setup and Field Equations}

We now describe our basic assumptions and main equations that we follow below.
In this work we consider a time-dependent and spherically-symmetric solution of Einstein field equations Eq.(\ref{Einstein})
in the presence of perfect fluid Eq.(\ref{PFEMT}). We adopt the isotropic coordinates and consider, without loss of generality,
the following line element
\begin{equation}
       ds^{2}=-T(t,r)dt^{2}+R(t,r)\left( dr^{2}+r^{2}d\Omega ^{2}\right).
\label{GenLineElement}
\end{equation}
where $T(t,r)$ and $R(t,r)$ are functions of the time $t$ and the radial coordinate $r$.
Introducing a dimensionless variable $x$ and the characteristic mass scale $m$ via
\begin{equation}
         r=\frac{m}{2}\left(1+x\right)
\label{lambda}
\end{equation}
we may rewrite the line element Eq.(\ref{GenLineElement}) as
\begin{equation}
         ds^{2}=-T(t,x)dt^{2}+\frac{m^{2}R(t,x)}{4}\left( dx^{2}+\left(
         1+x\right) ^{2}d\Omega ^{2}\right)
\label{lineelementTR}
\end{equation}
where for simplicity we still follow the same notation for the metric components as in Eq.(\ref{GenLineElement}).
Recognizing the fact that near Schwarzschild black hole event horizon the metric
acquires a typical $\textit{Rindler}\times \text{S}^{2}$ structure, we assume that as we approach towards the surface
$x=0$ the metric tends to
\begin{equation}
          \operatorname{ds^{2}}_{x \to 0} \simeq-\frac{x^{2}}{4}dt^{2}+4m^{2}\left(dx^{2}+
          d\Omega^{2}\right),
\label{Schw2}
\end{equation}
thereby capturing the causal structure associated with the corresponding
Rindler line element in the close vicinity to the $x=0$ event horizon.
We can easily convince ourselves that the constant $m$ coincides with
the mass in the Schwarzschild solution. The latter is obtained by replacing $a(t)$
in Eq.(\ref{McVittie}) with unity and tends to Eq.(\ref{Schw2}) in the $x \to 0$ limit.
As we see below, in a more general cases the constant $m$ does not acquire such simple meaning.

The line element Eq.(\ref{Schw2}) has no conical defects in its Euclidean regime,
once we have assumed that its Euclidean time coordinate has the correct periodicity $ \Delta\tau$
given by
\begin{equation}
          \Delta\tau=8 \pi m.
\label{periodicity}
\end{equation}
 Formally, the metric given by Eq.(\ref{Schw2}) serves as a boundary condition and also as
the zeroth order in our series solution
\begin{equation}
\begin{split}
       T(t,x)&=\frac{x^{2}}{4}(1+xf_{1}(t)+x^{2}f_{2}(t)+\dots)
\\
       R(t,x)&=16 \left( 1+xg_{1}(t)+x^{2}g_{2}(t)+\dots \right),
\end{split}
\label{metricTR}
\end{equation}
where $f_{n}(t), g_{n}(t)$ ($n=1,2,\dots$) are yet undetermined functions (to be referred
below as harmonics). An expansion around black hole horizon, in a non-cosmological context,
has been performed in \cite{Krishnan}.
Interestingly, as we show below, under the assumption that the line element Eq.(\ref{Schw2})
on the horizon has no conic singularity, i.e. satisfies Eq.(\ref{periodicity}),
the subsequent terms in the expansion must respect no conic singularity as well.
In other words, assuming Eq.(\ref{periodicity}) for $n=0$ terms is sufficient
to make all other $n>0$ terms periodic in the Euclidean time coordinate.

We now derive our main equations given by Eq.(\ref{PFEqState}) and Eq.(\ref{PFconstraint}) below,
utilized to obtain the analytic expressions for the functions $f_{n}(t), g_{n}(t)$ which were already defined in the metric
expansion Eq.(\ref{metricTR}). The obtained solution corresponds to a time-dependent
and spherically symmetric radial flow of a perfect fluid, such that the four velocity vector
has a non-vanishing radial component $U^{r} \neq 0$, and may be parametrized
according to
\begin{equation}
           U^{\mu}=\left(-\frac{\cosh(\varphi(t,r))}{\sqrt{T(t,r)}},\frac{\sinh(\varphi(t,r))}{\sqrt{R(t,r)}},0,0\right)
\label{4velocity}
\end{equation}
once the rapidity $\varphi(t,r)$ has been introduced.
Plugging the latter into Eq.(\ref{PFEMT})
one obtains an explicit expressions for the corresponding components of the perfect fluid
energy-momentum tensor, expressed through the following useful combinations $A,B,C,D$, defined by
\begin{equation}
\begin{split}
        &A=-2\mathcal{G}_{t}^{t}=\mathcal{T}_{t}^{t}=\rho \cosh ^{2}(\varphi (t,r))+P\sinh ^{2}(\varphi (t,r))
\\
        &B=2\mathcal{G}_{r}^{r}=-\mathcal{T}_{r}^{r}=\rho \sinh ^{2}(\varphi (t,r))+P\cosh ^{2}(\varphi (t,r))
\\
        &C=2\mathcal{G}_{\theta}^{\theta}=2\mathcal{G}_{\phi }^{\phi}
        =-\mathcal{T}_{\theta}^{\theta}=-\mathcal{T}_{\phi }^{\phi}=P
\\
        &D=-2\sqrt{\frac{\displaystyle T(t,r)}{\displaystyle R(t,r)}}\mathcal{G}_{r}^{t}=\sqrt{\frac{\displaystyle T(t,r)}{\displaystyle R(t,r)}}\mathcal{T}_{r}^{t}=
\\
         &\left( \rho +P\right) \sinh (\varphi(t,r))\cosh (\varphi (t,r)).
\end{split}
\label{ABCD}
\end{equation}
Here, we also utilized Einstein equations, i.e., proportionality between the Einstein tensor $\mathcal{G}^{\mu}_{\nu}$ and the
energy-momentum tensor $\mathcal{T}^{\mu}_{\nu}$.
The specific algebraic structure of the perfect fluid energy momentum tensor, implies that the emerging combinations $A,B,C,D$
are not independent. In fact, they are related by
\begin{equation}
            AB-D^{2}=C(A-B+C).
\label{isotropic1}
\end{equation}
Now, assuming the spacetime metric admits the form given by Eq.(\ref{lineelementTR}), one views Eq.(\ref{isotropic1}), as an equation
for the metric components $T(t,x)$ and $R(t,x)$. Formally speaking, by eliminating $\rho$, $P$, and $\varphi$ we have inverted the equations Eq.(\ref{ABCD})
and then by using the Einstein equations we ended up with the so-called perfect fluid isotropic condition.
This condition was originally obtained by McVittie and Wiltshire \cite{Walker},
and expresses the necessary condition on the metric to describe spherically symmetric flow of a perfect fluid.
Other works where time dependent spherically symmetric configuration have been studied may be found at \cite{McVittie2}, \cite{Ray}, \cite{BonnorKnusten}.
We should mention that in case we insist on the comoving solutions, which in principle are always possible for a single-component
perfect fluid,  Eq.(\ref{isotropic1}) is solved for $\mathcal{T}_{r}^{t}=0$ and
$\mathcal{T}_{r}^{r}=\mathcal{T}_{\theta}^{\theta}$  which lead to $U^{r}=0$.
In this work, however, we consider solutions which are easier to obtain and study in non-comoving  coordinates.
Numerous works, including the LTB dust solution \cite{LTB}, where the comoving coordinates are considered
may be found in the literature.

The perfect fluid isotropic condition is supplemented by matter equation of state, which in this work is chosen as the linear barotropic
\begin{equation}
        P=w\rho
\label{EqState}
\end{equation}
where $w$ is  some constant usually bounded between plus and minus unity.
By utilizing the combinations that were defined in
Eq.(\ref{ABCD}), the equation of state Eq.(\ref{EqState}) may be rewritten as
\begin{equation}
          C=w(A-B+C),
\label{PFEqState}
\end{equation}
while the isotropic condition Eq.(\ref{isotropic1}) acquires a simpler form
\begin{equation}
          AB-D^{2}=\frac{\displaystyle C^{2}}{\displaystyle w},
\label{PFconstraint}
\end{equation}
for $w \neq 0$. Clearly, the $w =0$ case should be tackled
with Eq.(\ref{isotropic1}).
Together the equations (\ref{PFEqState}), (\ref{PFconstraint}) constitute a system of two
basic coupled equations for the two unknown functions $T(t,x)$ and $R\left(t,x\right)$
which describe radial perfect fluid flow subject to the corresponding equation of state.
Those equations are difficult to solve even numerically and in the remainder of the paper we
will solve them perturbatively order by order around the local horizon,
according to the scheme mentioned near Eq.(\ref{metricTR}).

Finally, let us consider the underlying dynamics of the perfect fluid matter, which
is governed by the inherent
Bianchi identity and the corresponding energy-momentum conservation equation
\begin{equation}
       \nabla_{\mu }\mathcal{T}^{\mu \nu}=0.
       \label{EMTconserv}
\end{equation}
In case of perfect fluid energy-momentum tensor Eq.(\ref{PFEMT}) the latter
may be re-written as follows
\begin{equation}
\begin{split}
      \left( \rho +P\right)\left(U^{\nu} \nabla_{\mu } U^{\mu}+ U^{\mu} \nabla_{\mu }U^{\nu}\right) \quad
\\
      +\nabla_{\mu }\left( \rho +P\right) U^{\mu }U^{\nu } +\nabla_{\mu }Pg^{\mu \nu }=0.
       \label{EMTconservExplicit}
\end{split}
\end{equation}
The particular form of the energy-momentum tensor Eq.(\ref{PFEMT}), implies that
its divergence, given explicitly by Eq.(\ref{EMTconservExplicit}) is a sum of two terms.
Specifically, one term is parallel to the four velocity vector while the other is normal to it.
In fact, projecting Eq.(\ref{EMTconservExplicit}) on $U_{\nu}$ we find the component
that is parallel to the vector $U^{\nu}$
\begin{equation}
         \nabla_{\mu } \left( \rho U^{\mu} \right)+ P \nabla_{\mu} U^{\mu}=0,
\label{PreBConservation}
\end{equation}
which is known as the baryon number conservation equation.
The other term, known as the relativistic Euler equation, is orthogonal to the four-velocity
and is explicitly given by
\begin{equation}
           \left( \rho +P\right) U^{\mu} \nabla_{\mu }U^{\nu} \quad
\\
           +\nabla_{\mu }P\left( U^{\mu }U^{\nu } + g^{\mu \nu }\right)=0.
\label{Euler}
\end{equation}
Let us notice that Eq.(\ref{PreBConservation}) can be written as a vanishing divergence
of the current $j^{\mu}$
\begin{equation}
       \nabla_{\mu }j^{\mu}=0,
       \label{BNconserv}
\end{equation}
once we have introduced the proper density $n$ (number of particles per volume)
\begin{equation}
       j^{\mu}=n U^{\mu}
       \label{current}
\end{equation}
according to
\begin{equation}
         n=e^{\int \frac{d\rho }{\rho +P\left( \rho \right) }}.
\label{Nparticles}
\end{equation}
In the particular case of a linear barotropic equation of state given by
Eq.(\ref{EqState}), the proper density $n$ acquires the following simpler form
\begin{equation}
      n=\rho^{\frac{1}{1+w}}.
      \label{Nparticles}
\end{equation}
As a final remark we note that once the metric and the corresponding Einstein tensor have been worked out,
the following expressions for the rapidity $\varphi(t,r)$
\begin{equation}
\begin{split}
         \sinh^{2}(\varphi(t,r))=\frac{\displaystyle B-C}{\displaystyle A-B+2C}
\\
         \cosh^{2}(\varphi(t,r))=\frac{\displaystyle A+C}{\displaystyle A-B+2C}.
\end{split}
\label{4velABCD}
\end{equation}
may be utilized to determine the perfect fluid radial flow.


\section{No-go theorem for the static case}

We now prove that the only perfect fluid $P=w\rho$ static, spherically symmetric black hole solution is the Schwarzschild solution
with vanishing $P$ and $\rho$. For simplicity, we discuss first the $w=0$ case and turn to $w \neq 0$ case afterwards.
Then, we also consider the static case limit which proves to be useful when compared to the time-dependent case
in the following.

\subsection{The $w=0$ case}

Let us first consider the simpler static case with a vanishing proper pressure $P=0$, which
corresponds to the $w=0$ case. Assuming the spherically symmetric line element given by Eq.(\ref{GenLineElement})
let us define for convenience the functions $f(r), g(r)$
\begin{equation}
 T(r)=e^{f(r)} \,\,\, ,\,\,\, R(r)=e^{g(r)}
\end{equation}
and their derivatives with respect to the radial marker $r$
\begin{equation}
f^\prime(r)=F(r) \,\,\, ,\,\,\, g^\prime(r)=G(r).
\label{defFG}
\end{equation}
Substituting now the corresponding metric components into our basic equations Eq.(\ref{isotropic1}), Eq.(\ref{PFEqState})
we are driven towards the following first order differential equations for $F(r)$ and $G(r)$
\begin{equation}
\begin{split}
          2F(2+rF)+2(G+r(F^{\prime} + G^{\prime}))=0
\\
          2F(2+r G)+G(4+rG)=0. \quad
\label{EqForSchw}
\end{split}
\end{equation}
Solving those gives rise to
\begin{equation}
\begin{split}
         &F(r)=\frac{8m}{4r^{2}-m^{2}}
\\
        &G(r)=-\frac{4m}{r(m+2r)}
\end{split}
\end{equation}
and the corresponding Schwarzschild line element

\begin{equation}
          ds^{2}_{Sch}=-\left( \frac{1-\frac{\displaystyle m}{\displaystyle 2r}}{1+\frac{\displaystyle m}{\displaystyle 2r}} \right)^{2}dt^{2}
         +\left(1+\frac{m}{2r}\right)^{4}(dr^{2}+r^{2}d\Omega^{2}).
\label{Schw}
\end{equation}
This proves that the only static, spherically symmetric solution with dust perfect fluid is the dustless Schwarzschild solution.

Notice that in this simple case the proof did not rely on the assumption that local Killing horizon exists.
In the  following $w \neq 0$ case, however, we are not able to prove our theorem without
assuming the existence of a local Killing horizon.

\subsection{The $w \neq 0$ case}

Now let us consider the more interesting  $w \neq 0$ static case, with a non-vanishing pressure.
In such case our basic equations Eq.(\ref{PFEqState}), Eq.(\ref{PFconstraint}) are more involved,
and we are not able to solve our basic equations analytically. Nevertheless, assuming the existence of a local
Killing horizon at $x=0$, dictates $\textit{Rindler}\times\text{S}^{2}$ metric given by Eq.(\ref{Schw2}) as $x \to 0$.
This fixes the zeroth order terms in the expansion Eq.(\ref{metricTR}) and proves to be useful for a construction
of our solution as a power series in $x$, around $x=0$. Specifically, we assume that the functions $F(r), G(r)$
defined above, admit the following expansion
\begin{equation}
\begin{split}
           F&=\frac{\displaystyle \alpha_{-1}}{\displaystyle x}+\alpha_{0}+\alpha_{1}x+\alpha_{2}x^{2}+\ldots
\\
           G&=\beta_{0}+\beta_{1}x+\beta_{2}x^{2}+\ldots
\end{split}
\label{StaticCoeffEq}
\end{equation}
where $\alpha_{n-1}$ and $\beta_{n}$ ($n=0,1,2, \dots$) are the corresponding constants that we find order by order
by plugging the series Eq.(\ref{StaticCoeffEq}) into our basic equations Eq.(\ref{PFEqState}), Eq.(\ref{PFconstraint}).
The latter may be written as
\begin{equation}
\begin{split}
         P_{(1)}(r)+wP_{(2)}(r)=0
\\
          P_{(1)}^{2}(r)+wP_{(3)}(r)=0,
\end{split}
\label{P1P2P3}
\end{equation}
respectively, where $ P_{(1)}(r), P_{(2)}(r), P_{(3)}(r)$
 are $w$ independent quantities that are defined according to
\begin{equation}
\begin{split}
          & P_{(1)}(r)=(2+rF)F+2(G+r( F^\prime+ G^\prime))
\\
          & P_{(2)}(r)=(2-rF)F+10G+2r(FG+G^{2}- F^\prime)+G^\prime
\\
          & P_{(3)}(r)=(4(F+G)+2rFG+rG^{2})(8G+rG^2+4rG^\prime),
\end{split}
\end{equation}
where the prime denotes derivative with respect to $r$.
We can straightforwardly show that the previous solution in the $w=0$ case,
is still a solution of Eq.(\ref{P1P2P3}) for any  $w \neq 0$.
Furthermore, using the small $x$ expansion we can verify that in fact
this is the only solution, implying that $P_{(1)}(r)$, $P_{(2)}(r)$ and $P_{(3)}(r)$
all vanish.
Having solved Eq.(\ref{P1P2P3}) for the lowest order, i.e. $\alpha_{-1}$ and $\beta_{0}$, we proceed to the next, and so on
solving each time for the leading terms.
Following this procedure we are able to solve for all the following orders, and we bring
here some of the terms
\begin{equation}
\begin{split}
\displaystyle
          \alpha_{-1}&=\frac{1}{m}; \alpha_{0}=-\frac{1}{2m}; \alpha_{1}=\frac{1}{4m}; \alpha_{2}=\frac{2}{m}
\\
          \beta_{0}&=-\frac{1}{m}; \beta_{1}=\frac{3}{2m}; \beta_{2}=-\frac{7}{4m}.
\label{StatCoeff}
\end{split}
\end{equation}
From this expansion we can learn that it is $w$ independent and thus describes a Schwarzschild black hole.
To summarize, we have proven that any static metric with a Killing horizon in the presence of a perfect fluid, is necessarily
a Schwarzschild solution with a vanishing proper energy density and a vanishing proper pressure.
Interestingly, in the $w =0$ case, considered before, this claim holds even if we do not assume
the Killing horizon exists at $x=0$.

\subsection{Perfect Fluid in a Static Black Hole Background}

Now let us consider the perfect fluid flow in the static case limit \cite{Weinberg}, which provides some useful insights into the time-dependent case
that follows afterwards. In the static case the relativistic Euler equation Eq.(\ref{Euler})
is characterized by the time independent metric components $T(t,x)=T(x)$ and $R(t,x)=R(x)$
and by the fact that the fluid is static. This dictates that the four-velocity vector $U^{\mu}$ is purely timelike
\begin{equation}
            U^{0}=(-g_{00})^{-\frac{1}{2}} \qquad U^{\lambda} = 0 \quad \textrm{for} \quad \lambda \neq 0
\label{staticU}
\end{equation}
and also leads to the following identity

\begin{equation}
            \Gamma^{\mu}_{00}=-\frac{1}{2}g^{\mu\nu}\frac{\partial g_{00}}{\partial x^{\nu}}.
\label{staticC}
\end{equation}
Utilizing now Eq.(\ref{staticU}) and Eq.(\ref{staticC}) one may rewrite the relativistic Euler equation
given by Eq.(\ref{Euler}), for the space components $\nu=i$ ($i=1,2,3$), as following
\begin{equation}
           P_{,i}=-\frac{\rho+P}{T}T_{,i}
\label{Pdiv}
\end{equation}
(for $\nu=0$ the relativistic Euler equation is satisfied trivially in the static case).
The latter is nothing but the non-relativistic equation for hydrostatic equilibrium
in static gravitational field and has few interesting implications. In particular, assuming that such a spacetime is
also inhabited by a static black hole, then its event horizon (which also coincides with its
 Killing horizon in this static case) is identified by $T\vert _{r_{h}}=0$ where $r_{h}$ stands for
the horizon's coordinate. We notice now that under the assumption that $  (\rho+P)\vert _{r_{h}} \neq 0$
and finite $T_{,i}\vert _{r_{h}}$ on the horizon $r_{h}$, Eq.(\ref{Pdiv}) implies that the pressure gradient is necessarily divergent.
Therefore, avoiding this unphysical divergence calls for a vanishing of the following combination
\begin{equation}
          (\rho+P)\vert _{r_{h}} = 0
\label{rhoplusP}
\end{equation}
on the horizon $r_{h}$.
In case we also insist on the equation of state Eq.(\ref{EqState}), then we are also driven
towards a vanishing proper energy density on the horizon
\begin{equation}
        \rho\vert _{r_{h}}=0.
\label{rho0}
\end{equation}
As we will see below, the severe restriction given by Eq.(\ref{rhoplusP}) is elegantly removed once the
time dependence is introduced.

\section{Dynamics: Harmonic Expansion}

\subsection{First Harmonic}

We now turn to discuss the time-dependent solutions.
As we have mentioned above we hereby assume that our spacetime hosts a local Killing horizon, implying that as
$x \rightarrow 0$ the components of our metric Eq.(\ref{lineelementTR})  admit the
expansion given by Eq.(\ref{metricTR}). In particular, in the leading order the corresponding line element
is given by Eq.(\ref{Schw2}).
The governing equations for the first harmonics $f_{1}(t)$ and $g_{1}(t)$, are found once we plug
the expansion  Eq.(\ref{metricTR}) into our basic equations Eq.(\ref{PFEqState}), Eq.(\ref{PFconstraint})  and keep the leading order in
each one accordingly.
This way, we obtain the following equations for the first harmonics
\begin{equation}
      1+w+\left( 1-w\right) f_{1}\left( t\right) +wg_{1}\left( t\right) -16m^{2}g_{1}^{\prime \prime }\left( t\right) =0
\end{equation}
\begin{equation}
       1+f_{1}\left( t\right) -16m^{2}g_{1}^{\prime \prime }\left(
       t\right) =0.
\end{equation}
which are valid for any choice of the parameter $w$.
The latter are linear and admit an analytical solution explicitly given by
\begin{equation}
\begin{split}
           g_{1}(t)&=-1+pe^{\omega t}+qe^{-\omega t}
\\
           f_{1}\left( t\right)&=-2+pe^{\omega t}+qe^{-\omega t}
\end{split}
\label{FirstH}
\end{equation}
where the Euclidean angular frequency $\omega$ has been introduced according to
\begin{equation}
         \omega=1/4m
\end{equation}
Here we notice, that in order that our expansion stays valid it is actually needed
that the following product remains small
\begin{equation}
         xe^{\pm \omega t} \ll 1.
\end{equation}
From the first order terms Eq.(\ref{FirstH}) given above, we can already learn about one of the major features
of our time-dependent solution which holds to higher harmonics as well. Particularly, it has no conical singularity once continued
to the Euclidean time via $t=-it_{E}$. In fact, keeping in mind that the correct
periodicity Eq.(\ref{periodicity}) of the Euclidean metric Eq.(\ref{Schw2}) is assumed to hold,
we can check by substitution that a shift $\Delta \tau$,
\begin{equation}
           \pm i \omega (t_{E}+\Delta\tau)= \pm i \omega t_{E}\pm 2\pi i
\end{equation}
gives rise to the $2\pi i$ shift.

Dropping the time-dependence
by considering  $p =q= 0$ case simply brings us back to the Schwarzschild,
and only  $p\not=0$,  $q\not=0$ can go beyond.
In fact, calculating the proper energy density $\rho$ on the horizon up to
the first order, yields
\begin{equation}
      \rho =-\frac{24 \omega^{2}pq(1-3w)}{1+w}+O(x)
\label{rhofirst}
\end{equation}
that depends on the product $pq$. This just reflects the fact, that our
solution describes a static horizon, and therefore any shift of the time
coordinate $t\rightarrow t+\Delta t$ results in
$p\rightarrow pe^{\omega \Delta t}$, $q\rightarrow qe^{-\omega \Delta t
}$ so that the product $pq \rightarrow pq$ stays unchanged.
The non-vanishing proper energy density on the horizon, in this time-dependent
case is in contrast with our previous comment near Eq.(\ref{rho0}), where we
have indicated that the proper density $\rho$ should vanish in a static case.

Interestingly, the first harmonic may be fixed by an alternative demand that $\rho$ and $P$
are non-singular on the horizon, without providing any particular equation of state.
To this end, let us calculate to the leading order the proper energy density, the pressure,
and the perfect fluid isotropic condition Eq.(\ref{isotropic1})
\begin{equation}
          \rho=\frac{4\omega^{2}\left(-1+ f_{1}(t)-g_{1}(t)\right)}{x}+O(x^{0})
\label{rho1overx}
\end{equation}
\begin{equation}
          P=\frac{4\omega^{2}\left(1+f_{1}\left( t\right) -\frac{1}{\omega^{2}}g_{1}^{\prime \prime }(t)\right)}   {x}+O(x^{0})
\label{P1overx}
\end{equation}
\begin{equation}
\begin{split}
         &AB-D^{2}-C(A-B+C)
\\
        &=-\frac{16\omega^{4}\left(-1+f_{1}(t)-g_{1}(t)\right)\left(1+f_{1}\left( t\right) -\frac{1}{\omega^{2}}g_{1}^{\prime \prime }(t)\right)}{x^{2}}
\\
        &+O(x^{-1}),
\end{split}
\end{equation}
respectively. Assuming now the leading terms in Eq.(\ref{rho1overx}) and
Eq.(\ref{P1overx}) vanish, we are guaranteed that perfect fluid isotropic
condition is satisfied.
The perfect fluid is non static by construction, see Eq.(\ref{4velocity}).
Furthermore, Eq.(\ref{4velABCD}) then tells us that associated with our solution is
the rapidity
\begin{equation}
          \sinh (2\varphi \left( t,r\right) )=
          \frac{p^{2}e^{2\omega t}-q^{2}e^{-2\omega t}}  {2pq}+{\cal O}(x) ~.
\label{rapidityfirst}
\end{equation}
The important point is not that $T_{\mu\nu}$ is non-diagonal, but
that there is a radial current or flow represented by this rapidity.

But this is not all there is; the physical quantities such as proper energy density $\rho$ are related to the Ricci tensor, which involves second order metric derivatives. Therefore, it seems that we must
go even further, at least to the second harmonic for additional contribution.

\subsection{Second harmonic}

Before we consider the full set of equations for the second order terms $f_{2}(t)$ and $g_{2}(t)$,
let us try to learn how the restriction Eq.(\ref{rhoplusP}) is removed once
the time-dependence is introduced.
To this end, we consider the expansion Eq.(\ref{metricTR}) up to the second order,
with the first harmonic Eq.(\ref{FirstH}) plugged in.
This way, we can learn about the role of just the perfect fluid constraint Eq.(\ref{PFconstraint}),
but still without specifying an equation of state. Of particular interest
are the proper energy density, and the proper pressure that may be expressed up to
the second order terms, according to
\begin{equation}
          \rho=12\omega^{2}\Big(\frac{11}{4}-pe^{\omega t}-qe^{-\omega t}-pq
          +f_{2}(t)-g_{2}(t)\Big)
\label{rho2}
\end{equation}
\begin{equation}
\begin{split}
          P=4\omega^{2}\Big( -\frac{23}{4}&+\frac{9p}{2}e^{\omega t}+\frac{9q}{2}e^{-\omega t}-
\\
         &3pq+ 3f_{2}(t)+g_{2}(t)-\frac{1}{\omega^{2}} g_{2}^{\prime\prime}(t)\Big).
\end{split}
\label{P2}
\end{equation}
Keeping in mind that the sum $\rho+P$ is significant in the static case,
we bring here the corresponding expression up to the second order
\begin{equation}
\begin{split}
           \rho+P=2\omega^{2}\Big(&5+3pe^{\omega t}+3qe^{-\omega t}-12pq+
\\
            &12f_{2}(t)-4g_{2}(t)-\frac{2}{\omega^{2}} g_{2}^{\prime\prime}(t)\Big)
\end{split}
\label{rhoplusP2}
\end{equation}
and also the static case limit of the expressions given by Eq.(\ref{rho2}), Eq.(\ref{P2}), Eq.(\ref{rhoplusP2})
\begin{equation}
\begin{split}
          &\rho=3\omega^{2}\left(11+4f_{2}-4g_{2}\right)
\\
          &P=\omega^{2}\left( -23+12f_{2}+4g_{2}\right)
\\
          &\rho+P=2\omega^{2}\left( 5+12f_{2}-4g_{2}\right).
\end{split}
\label{StaticRhoPSecOrder}
\end{equation}
Interestingly, there is a close relation between the perfect fluid isotropic condition Eq.(\ref{PFconstraint})
and the vanishing of the mentioned combination $\rho+P = 0$.
Specifically, as may be verified with the help of MathTensor package (some of the expressions are too long to be written here explicitly)
the perfect fluid isotropic condition is equivalent to the condition $(\rho+P)\vert _{r_{h}} = 0$ in the static case.
In such case characterized by $f_{2}(t)=f_{2}$ and $g_{2}(t)=g_{2}$, where $f_{2}$
and $g_{2}$ are some constants, those two conditions imply
\begin{equation}
 g_{2}=\frac{5}{4}+3f_{2} \,\,\, , \,\,\, \rho=-P=6\omega^{2}(3-4f_{2}),
\end{equation}
still without specifying a concrete equation of state.
An attempt to impose on top of it an equation of state of the type $P=w \rho$,
will take us back to Eq.(\ref{StatCoeff}) and to the corresponding coefficients
given $f_{2}=\frac{3}{4}$ and $g_{2}=\frac{7}{2}$, which brings us back to
Eq.(\ref{rho0}).

Once $t-$dependence is introduced, imposing $\rho+P=0$ leaves us with a
full differential equation for $g_{2}(t)$. Had we imposed also $P= w \rho$
we would have faced another full differential equation.
In fact, the relevant equations for $g_{2}(t)$ and $f_{2}(t)$ are obtained by substituting the
near horizon expansion Eq.(\ref{metricTR}) with already known first order terms Eq.(\ref{FirstH}),
into our basic equations Eq.(\ref{PFEqState}), Eq.(\ref{PFconstraint}).
Keeping the leading terms up to the second order we derive the equations
for the corresponding $f_{2}(t)$ and $g_{2}(t)$. Quite generally, Eq.(\ref{PFEqState})
does not comprise any derivatives of the function $f_{2}(t)$, which allows
to express the latter through $g_{2}(t)$ and its derivatives according to
\begin{equation}
\begin{split}
          f_{2}(t)&=\frac{1}{12(-1+w)}\Big(-23-33\omega+6p(3+2w)e^{\omega t}
\\
         &+6q(3+2w)e^{-\omega t}+12pq(-1+w)
\\
         &+4(1+3w)g_{2}(t)-\frac{4}{\omega^{2}}g^{\prime\prime}_{2}(t)\Big)
\end{split}
\label{f2fromg2}
\end{equation}
Plugging this relation
into the corresponding leading order of the perfect fluid isotropic condition Eq.(\ref{PFconstraint}),
we are led towards the following differential equation for $g_{2}(t)$
\begin{equation}
\begin{split}
          -w s^{\prime\prime}(t)^{2}+2\omega^{2}(1+w)^{2}s(t)s^{\prime\prime}(t)-
\\
          \omega^{2}(1-w)^{2}s^{\prime}(t)^{2}-4\omega^{4}(1+w)^{2}s(t)^{2}=0
\end{split}
\label{s}
\end{equation}
where
\begin{equation}
\begin{split}
           s(t)=&14-10\left(p e^{\omega t}+q e^{-\omega t}\right)+
\\
          &3\left(p^{2} e^{2\omega t}+q^{2} e^{-2\omega t}\right)-4g_{2}(t).
\end{split}
\label{def_s}
\end{equation}
In a particular case of a dust perfect fluid, associated with $w=0$, the latter can be
furthermore simplified
\begin{equation}
         \frac{d^{2}}{dt^{2}}\sqrt{s(t)}=\omega^{2}\sqrt{s(t)}.
\end{equation}
The solution of Eq.(\ref{s}) is evidently given by
\begin{equation}
           s(t)=-4\alpha_{0}\left(2(1-w)+(1+w)^{2}e^{2\omega (t-\beta_{0})}+e^{-2\omega(t-\beta_{0})}\right).
\end{equation}
which according to Eq.(\ref{f2fromg2}), Eq.(\ref{def_s}) yields an explicit expressions for the second order terms
\begin{equation}
\begin{split}
          f_{2}(t)&=\frac{3}{4}+pq\Big(1-\xi \frac{1+3w}{2(1+w)}\Big)-\frac{3}{2}\Big(p e^{\omega t}+q e^{-\omega t} \Big)
\\
         &+\frac{3}{4}k\Big(\frac{p^{2}}{\eta}e^{2\omega t}+\frac{q^{2}}{1-\eta}e^{-2\omega t} \Big)
\\
          g_{2}(t)&=\frac{7}{2}+\xi pq \frac{3(1-w)}{2(1+w)}-\frac{5}{2}\Big(p e^{\omega t}+q e^{-\omega t} \Big)
\\
         &+\frac{3}{4}k\Big(\frac{p^{2}}{\eta}e^{2\omega t}+\frac{q^{2}}{1-\eta}e^{-2\omega t} \Big),
\end{split}
\label{SecondH}
\end{equation}
where the constants $k$, $\eta$, and $\xi$ have been introduced according to
\begin{equation}
\begin{split}
          2\alpha_{0}&=\frac{3}{2}\xi pq
\\
          \frac{4\alpha_{0}}{3}e^{-2\omega \beta_{0}}&=\left(\frac{k}{\eta}-1 \right)^{2} p^{2}
\\
           \frac{4\alpha_{0}}{3}e^{2\omega \beta_{0}}&=\left(\frac{k}{1-\eta}-1 \right)^{2}q^{2}.
\end{split}
\end{equation}
The coefficients $k,\eta , \xi$ are not independent and in fact, are found to respect the following consistency relation
\begin{equation}
       \xi ^{2}=1-\frac{\left( 1-k\right) k}{\left( 1-\eta \right) \eta }.
\label{consistency}
\end{equation}

We now discuss the physical quantities, such as the energy density and the
rapidity associated with our solution.
In fact, the latter is non-trivial since we allow an accreting solution with non-vanishing
radial component of the four-velocity vector, as specified by Eq.(\ref{4velocity}).
The proper energy density on the horizon and the rapidity are given now by
\begin{equation}
      \rho =-\frac{24\omega^{2}\xi pq}{1+w}+O(x)
\label{rhosecond}
\end{equation}
\begin{equation}
\begin{split}
\displaystyle
             & \sinh (2\varphi \left( t,r\right) )=
\\
              &\frac{q^{2}\eta \left( 1-k-\eta \right) e^{-2\omega t}+p^{2}\left( \eta -k\right) \left( \eta -1\right)
               e^{2\omega t}}{2 \xi pq\eta \left( \eta -1\right)}+O\left( x\right),
\end{split}
\label{rapiditysecond}
\end{equation}
respectively.
As expected, since the proper density and the rapidity depend on the second derivatives of the metric,
the obtained relations Eq.(\ref{rhosecond}) and Eq.(\ref{rapiditysecond}) differ from the previously obtained expressions
Eq.(\ref{rhofirst}) and Eq.(\ref{rapidityfirst}).

From expressions  Eq.(\ref{rhosecond}) and Eq.(\ref{rapiditysecond}) we learn that:
(i) The local Killing horizon is non-singular at $x=0$ for any constant $w$ (except
$w=-1$). (ii) The local Killing horizon hosts non vanishing energy density. (iii) The combination
which appears to be of physical significance is $\xi pq$ and not just $pq$.
While (i) and (ii) are evident from Eq.(\ref{rhosecond}), (iii) has additional implications and deserves
further discussion.
In fact, as we demonstrate below, various scalar invariants are all proportional to $\xi pq$ implying also
that the physical quantities $\rho$ and $P$ are proportional to this combination as well.
This suggests that the expression $p e^{\omega t}+q e^{-\omega t}$
which appears in Eq.(\ref{FirstH}) can in fact be gauged away completely, i.e. $p,q \rightarrow 0$,
provided we insist $\xi pq \rightarrow const$.
Specifically, after taking this limit, the first order term given by Eq.(\ref{FirstH}) turns into

\begin{equation}
\begin{split}
         &f_{1}\left( t\right)=-1
\\
          &g_{1}\left( t\right) =-2.
\end{split}
\label{1stfinal}
\end{equation}

In the second harmonic, given by Eq.(\ref{SecondH}), taking the product $p,q\rightarrow 0$ but
insisting on keeping the product $\xi pq$ constant we are driven towards $\xi \sim \frac{1}{pq}$,
that  should agree with the consistency relation Eq.(\ref{consistency}) between $\xi $, $p$ and $q$.
Under the following redefinition
\begin{equation}
\begin{split}
       &\xi\rightarrow \frac{\xi }{pq};
\\
      &\frac{1-\eta }{\eta }\rightarrow \left(
      \frac{1-\eta }{\eta }\right) \frac{q^{2}}{p^{2}};
\\
      &\frac{k^{2}}{\eta \left(
      1-\eta \right) }\rightarrow \frac{k^{2}}{\eta \left( 1-\eta \right) }\frac{1}{p^{2}q^{2}}
\end{split}
\end{equation}
the redefined consistency relation in the $p,q\rightarrow 0$ limit reads
\begin{equation}
       \xi ^{2}=\frac{k^{2}}{\left( 1-\eta \right) \eta }
\end{equation}
for $0<\eta <1$. Furthermore, utilizing the freedom of choice of coordinates $t\rightarrow t+\Delta t$,
that does not change $\xi $, without loss of generality we can always choose
$\frac{k}{\eta }e^{2\omega \Delta t}=\frac{k}{1-\eta }e^{-2\omega \Delta t}$.
This way the second harmonic may acquire the following form
\begin{equation}
\begin{split}
           f_{2}\left( t\right) =\frac{3}{4}-\frac{1+3w}{2\left( 1+w\right) }\xi +\frac{3}{2}\xi \cosh \left(2\omega t\right)
\\
          g_{2}\left( t\right) =\frac{7}{2}+\frac{3\left( 1-w\right) }{2\left(1+w\right) }\xi +
          \frac{3}{2}\xi \cosh \left(2 \omega t\right)
\end{split}
\label{2ndfinal}
\end{equation}
while the rapidity is given by

\begin{equation}
            \sinh(2\varphi(t,r))= -\sinh\left(2\omega t\right)+O(x^{2}).
\end{equation}

\subsection{Third Harmonic and Curvature Scalars}

We may follow the general prescription to the third ($n=3$) order.
Specifically, let us substitute the metric components given by Eq.(\ref{metricTR}),
such that the only present terms are of order three and lower. Keeping in mind that the lower orders $n=1$ and
$n=2$ are already known and are given by Eq.(\ref{1stfinal}) and Eq.(\ref{2ndfinal}), respectively, we
expect to obtain equations for $f_{3}(t)$ and $g_{3}(t)$.
Similarly to what we have done in the lower orders, Eq.(\ref{PFEqState}) allows to express
$g_{3}(t)$ through $f_{3}(t)$ and its derivatives. Plugging the resulting expression for $g_{3}(t)$ into our second basic Eq.(\ref{PFconstraint})
we end up with a second order differential equation for $f_{3}(t)$.
Solving this equation for $f_{3}(t)$ yields

\begin{equation}
\begin{split}
          f_{3}(t)=&\left(-\frac{1}{2}+\frac{1+3w}{1+w}\xi\right)-
\\
          &\frac{1}{3}(1+3w)\Big(e^{\omega t}(2p_{3}(1+w)+q_{3}(1-w))+
\\
          & e^{-\omega t}(p_{3}(1-w)+2q_{3}(1+w))\Big)-
\\
         3\xi \cosh\left(2\omega t\right)&+\frac{1}{3}(3+w)(1+3w)\left(p_{3}e^{3\omega t}+q_{3}e^{-3\omega t}\right)
\end{split}
\label{thirdf3}
\end{equation}

\begin{equation}
\begin{split}
          g_{3}(t)=&-\left(\frac{11}{2}+\frac{9(1-w)}{2(1+w)}\xi\right)-
\\
          &(1-w)\Big(e^{\omega t}(2p_{3}(1+w)+q_{3}(1-w))+
\\
          & e^{-\omega t}(p_{3}(1-w)+2q_{3}(1+w))\Big)-
\\
           \frac{9}{2}\xi \cosh\left(2\omega t\right)&+\frac{1}{3}(3+w)(1+3w)\left(p_{3}e^{3\omega t}+q_{3}e^{-3\omega t}\right)
\end{split}
\label{thirdg3}
\end{equation}
where $p_{3}$ and $q_{3}$ are some constants.

\section{Higher Order Invariants, Baryon Number Conservation Equation,
General Structure}

\subsection{Higher Order Invariants}

Is the horizon surface at $x=0$ singular? To answer this question
one may calculate some curvature scalars and check their regularity on the horizon.
In fact, utilizing the MathTensor package  we can work out the following expressions for
the various curvature scalars $\mathcal{R}$,
$\mathcal{R^{\mu\nu}R_{\mu\nu}}$, $\mathcal{R^{\mu\nu\lambda\sigma}R_{\mu\nu\lambda\sigma}}$
along the metric Eq.(\ref{lineelementTR}),
\begin{equation}
            \mathcal{R}=\frac{24\omega^{2}\xi(1-3w) }{1+w}+32\omega^{2}(1-3w)W(t)x+O(x^{2})
\end{equation}
\begin{equation}
\begin{split}
            &\mathcal{R^{\mu\nu}R_{\mu\nu}}=\frac{576 \omega^{4}\xi^{2}(1+3w^{2}) }{(1+w)^{2}}+
            \frac{1536\omega^{4}\xi(1+3w^{2})}{1+w}W(t)x
\\
            &+O(x^{2})
\end{split}
\end{equation}
\begin{equation}
\begin{split}
          &\mathcal{R}^{\mu \nu \lambda \sigma}\mathcal{R}_{\mu \nu \lambda \sigma}=
\\
         &\frac{252\omega^{4}(4\xi(1+w)+(1+w)^{2}+\xi^{2}(9+6w+9w^{2}))}{(1+w)^{2}}
\\
         &+\frac{512 \omega^{4}(2(1+w)+\xi(9+6w+9w^{2}))}{1+w}W(t)x
         +O(x^{2})
\end{split}
\end{equation}
where $W(t)$ is given by

\begin{equation}
\begin{split}
              W(t)=&p_{3}\left(2(1+w)e^{\omega t}+(1-w)e^{-\omega t}\right)+
\\
             &q_{3} \left((1-w)e^{\omega t}+2(1+w) e^{-\omega t}\right).
\end{split}
\end{equation}
As expected, those higher order invariants are regular on the local Killing horizon $x=0$
and its close vicinity, signaling that there is no physical singularity in our solution.

\subsection{Baryon Number Conservation Equation}

The perfect fluid isotropic condition Eq.(\ref{PFEqState}) and the associated equation of state Eq.(\ref{PFconstraint})
form the basic equations for the metric components. The perfect fluid equations of motion, on the other hand,
are governed by the inherent energy-momentum conservation Eq.(\ref{EMTconserv}) and the four-velocity vector
normalization condition Eq.(\ref{4vel}). In fact, according to our discussion above those lead to the Euler and Baryon number
conservation equations, given by Eq.(\ref{Euler}), Eq.(\ref{BNconserv}), respectively.
Consequently, it is possible to perform a consistency check, by plugging our solution
into the right hand side of the baryon number conservation equation
\begin{equation}
          \frac{1}{\sqrt{-g}} \partial_{\mu}(\sqrt{-g}\rho^{\frac{1}{1+w}}U^{\mu})=0
\label{BN2}
\end{equation}
keeping in mind that it should vanish along our solution.
In our case of study, the line element and the four velocity vector are given by Eq.(\ref{GenLineElement})
and Eq.(\ref{4velocity}), respectively, and Eq.(\ref{BN2}) acquires the following form
\begin{equation}
          r^{2}\left(\sqrt{R}\rho^{\frac{1}{1+w}}\cosh(\varphi) \right)_{,t}+ \left(\sqrt{T}r^{2}\rho^{\frac{1}{1+w}}\sinh(\varphi) \right)_{,r}=0.
\end{equation}
Plugging in now the harmonics up to third order, we arrive towards the following
\begin{equation}
          4p^{2}q^{2}(-1+\eta)^{2}\eta^{2}\left(\xi^{2}-1+\frac{(-1+k)k}{(-1+\eta)\eta} \right)=0
\end{equation}
which holds thanks to Eq.(\ref{consistency}) we obtained before.
As expected, we verify that our solution respects the baryon number conservation equation
and that the latter imposes no further restrictions at the zeroth order.

\subsection{General Structure}

Let us assume that we have calculated the terms in our expansion Eq.(\ref{metricTR}) of order $n-1$ and lower.
Writing next the equation of state Eq.(\ref{PFEqState}) for $f_{n}(t)$ and $g_{n}(t)$ and analyzing it,
one can see it admits the following properties:
(i) doesn't comprise time derivatives of $f_{n}(t)$ and
(ii) linear with respect to $f_{n}(t)$.
Similarly to the expression Eq.(\ref{f2fromg2}) this allows to represent  $f_{n}(t)$
through $g_{n}(t)$ and its derivatives.
Finally substituting the latter into the other Eq.(\ref{PFconstraint}) one ends up with a linear equation for $g_{n}(t)$.
Carrying out this steps one may obtain the third order terms $f_{3}(t)$ and $g_{3}(t)$ in the expansion Eq.(\ref{metricTR}).
Those are given explicitly by Eq.(\ref{thirdf3}), Eq.(\ref{thirdg3}), respectively.

At each order $n$ of $x$ two linear differential equations for $f_{n}(t)$ and $g_{n}(t)$ are introduced.
This suggests the following double series structure
\begin{equation}
\begin{split}
           T(t,x)&=\frac{1}{4}x^{2} \sum_{n=0}^{\infty} f_{n}(t)x^{n}
\\
           R(t,x)&=16\sum_{n=0}^{\infty} g_{n}(t)x^{n}
\end{split}
\end{equation}
such that
\begin{equation}
\begin{split}
           f_{n}(t)&=\sum_{m=0}^{n}a_{m}^{(n)}e^{\pm m \omega t}
\\
           g_{n}(t)&=\sum_{m=0}^{n}b_{m}^{(n)}e^{\pm m \omega t}.
\end{split}
\end{equation}
The above structure can be put into the alternative form
\begin{equation}
\begin{split}
            T(t,x)&=\frac{1}{4}x^{2} \sum_{m=0}^{\infty} a_{m}(x)e^{\pm m \omega t}
\\
            R(t,x)&=\sum_{m=0}^{\infty} b_{m}(x)e^{\pm m \omega t}.
\end{split}
\end{equation}

Note that the explicit Euclidean $t$ periodicity has not really gone away; apart from the transformation itself, it enters now to the circumferential radius.

\section{Summary}

In this paper, we have proven the existence of general relativistic solutions describing black holes embedded in  accreting perfect fluid, and demonstrated the phenomenon
for the class of $P=w\rho$ equations of state.
Obviously, by simply failing to respect the assumed equation of state,
the well known McVittie solution does not fall into this category.
A perturbative analysis is performed, and explicit solutions for the
corresponding metric components have been obtained.
While counter intuitively admitting a local time-like Killing vector
on the event horizon itself, the various black hole configurations
are necessarily time dependent (thereby avoiding a well known
no-go theorem) away from the horizon.
Consistently, Hawking's imaginary time periodicity is globally
manifest on the spacetime manifold.

\bigskip

\acknowledgments{}

\end{document}